\documentclass[aps,prl,
twocolumn,
showpacs,superscriptaddress,groupedaddress,amsmath,amssymb]{revtex4}
\usepackage{graphicx}
\usepackage{dcolumn}
\usepackage{bm}

\begin{document}

\preprint{APS/123-QED}

\title{Shape Transitions in Hyperbolic Non-Euclidean Plates}

\author{John Gemmer}
 \email{jgemmer@math.arizona.edu}
 \affiliation{Arizona Center for Mathematical Sciences, University of Arizona, Tucson, Arizona, USA}

\author{Shankar Venkataramani}%
\affiliation{Department of Mathematics, University of Arizona, Tucson, Arizona USA}



\date{\today}

\begin{abstract}
We present and summarize the results of recent studies on non-Euclidean plates with imposed constant negative Gaussian curvature in both the F\"oppl - von K\'arm\'an  and Kirchhoff approximations. Motivated by experimental results we focus on annuli with a periodic profile. We show that in the F\"oppl - von K\'arm\'an approximation there are only two types of global minimizers -- flat and saddle shaped deformations with localized regions of stretching near the boundary of the annulus.  We also show that there exists local minimizers with $n$-waves that have regions of stretching near their lines of inflection. In the Kirchhoff approximation we show that there exist exact isometric immersions with periodic profiles. The number of waves in these configurations is set by the condition that the bending energy remains finite and grows approximately exponentially with the radius of the annulus. For large radii, these shape are energetically favorable over saddle shapes and could explain why wavy shapes are selected by crocheted models of the hyperbolic plane.
\end{abstract}

\maketitle


When subjected to localized lateral swelling or growth, thin elastic sheets can form periodic rippling patterns as observed in torn plastic sheets \cite{Sharon2002BuckleCascades}, along the edges of leaves \cite{HaiyiLiang12292009} and in swelling hydrogels \cite{Shankar2011Gels}. Since the bending energy of thin sheets is much weaker than stretching, the morphology of these structures is a result of the sheet bending to relieve in-plane strain. Recently, there has been interest in using swelling as a mechanism for shaping flat sheets into desired three-dimensional structures \cite{Shankar2011Gels, Santangelo12}. This type of specified morphogenesis has been experimentally realized by varying the radial shrinkage factor of thermally responsive gel disks \cite{Shankar2011Gels} and by directly controlling the local swelling at specific points through halftone gel lithography  \cite{Santangelo12}. 

The equilibrium configuration of such sheets can be modelled as the minimum of a non-Euclidean elastic energy that measures strains from a fixed two-dimensional Riemannian metric $\mathbf{g}$ defined on the midsurface of the sheet $\mathcal{D}\subset \mathbb{R}^2$ \cite{Efrati2009nonEucplates}. The intuition behind this model is that $\mathbf{g}$ encodes how the local swelling changes the intrinsic distance between material coordinates and the body deforms to be as ``close as possible'' to an isometric immersion of $\mathbf{g}$.  In this model, deformations with vanishing stretching energy correspond to isometric immersions of $\mathbf{g}$ into $\mathbb{R}^3$ and in the vanishing thickness limit a particular isometric immersion, provided one exists, is selected by the bending energy \cite{linearizedGeometry}.

This letter summarizes two studies of swelling thin elastic sheets that we presented in \cite{Gemmer2011} and \cite{Gemmer2011preprint}. In these works we considered the F\"oppl - von K\'arm\'an (FvK) and Kirchhoff approximations when $\mathcal{D}$ is an annulus with inner radius $r_0$ and outer radius $R$ and $\mathbf{g}$  has corresponding Gaussian curvature $K_0$ that is constant and negative. In geodesic polar coordinates $(\rho,\theta)$ the metric takes the form $\mathbf{g}=dr^2+|K_0|^{-1}\sinh^2(\sqrt{|K_0|})d\theta^2$ and isometric immersions of $\mathbf{g}$ correspond to (local) isometric immersions of the hyperbolic plane $\mathbb{H}^2$ \cite{gray}. 

Physical models of such sheets of have been constructed by crocheting disks such that the circumference of the disk grows exponentially with the radius \cite{Taimina-2001}. As the radius increases, the profile of these disks transition from saddle shaped to wavy with a refinement of the number of waves. Hydrogel disks with both radius and $K_0$ fixed have also been constructed \cite{Shankar2011Gels}. In these gels it was observed that the equilibrium configurations obtain a periodic profile of one wavelength with a refinement of the number of waves $n$ with decreasing thickness that is well fit by the power law $n\sim t^{-\frac{1}{2}}$ \cite{Shankar2011Gels}. Motivated by these results we focused on deformations with a periodic profile with $n$-fold odd rotational symmetry.

\begin{figure}
\begin{center}
\includegraphics[width=3.4in]{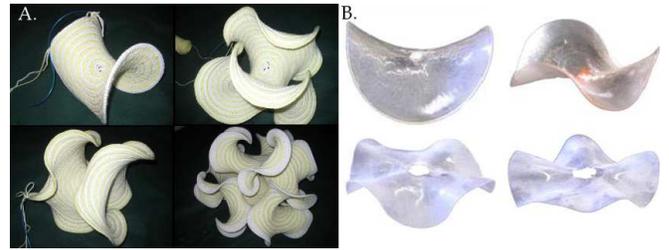}
\end{center} 
\caption{\textbf{A}: The evolution of the crocheted hyperbolic plane. As the size of the disk grows, increasing radius clockwise from upper left, the profile obtains a wavy pattern that becomes increasingly complex. Images courtesy of Gabriele Meyer. \textbf{B}: The equilibrium shapes of hydrogel disks of constant radius $R=14mm$, constant Gaussian curvature $K_0=-.0011mm^{-2}$, and varying thickness $t=.75mm$, $t=.6mm$, $t=.25mm$ and $t=.19mm$ clockwise from upper left \cite{Shankar2011Gels}. \label{fig:examples}}
\end{figure}

%
%
\emph{F\"oppl - von K\'arm\'an approximation} -- In the FvK approximation the deformation of the sheet is decomposed into in-plane $\epsilon^2\chi:\mathcal{D}\mapsto \mathbb{R}^2$ and out-of-plane displacements $\epsilon\eta:\mathcal{D}\mapsto \mathbb{R}$, where $\epsilon=\sqrt{|K_0|R}$ is a dimensionless curvature. Expanding in $\epsilon=\sqrt{|K_0|}R$ it follows that $\mathbf{g}\approx \mathbf{g}_0+\epsilon^2 \mathbf{g}_1$, where $\mathbf{g}_0$ is the Euclidean metric and $\mathbf{g}_1=\rho^4/(3R^2)\,d\theta^2$. Thus to lowest order  the in-plane strain tensor is $\gamma=(\nabla \chi)^T+\nabla \chi+(\nabla \eta)^T\cdot \nabla \eta-\mathbf{g}_1$, which retains the nonlinear contribution to the strain from the out-of-plane displacement. 

The energy normalized by $t$ can then be approximated by
\begin{eqnarray}
E_F[\chi,\eta]=\mathcal{S}[\gamma]+\tau^2\mathcal{B}[D^2\eta] \label{FvK:Energy},
\end{eqnarray}
where $\tau=t/(\sqrt{3}R\epsilon)$ is a dimensionless thickness, $D^2\eta$ denotes the Hessian matrix of second partial derivatives of $\eta$ and the energies $\mathcal{S}$ and $\mathcal{B}$ are the lowest order approximate to the stretching and bending energies in \cite{Efrati2009nonEucplates}.  Moreover, $D^2\eta$ is the lowest order approximation to the second fundamental form of the deformation and its invariants $\text{tr}(D^2\eta)=\Delta \eta$ and $\det(D^2\eta)$ are approximations to the mean and Gaussian curvatures respectively.


 There are two deformations, flat and stretching free,  that capture the scaling with $\tau$ of minimizers of $E_F$. Flat deformations correspond to deformations with $\eta=0$, satisfy $\mathcal{B}[\eta]=0$, and the global minimum $\mathcal{F}$ over such deformations can be found by solving a linear boundary value problem \cite{Gemmer2011preprint}.  Stretching free deformations correspond to deformations with $\gamma=0$, satisfy $\mathcal{S}[\gamma]=0$, and can be found by solving the Monge-Ampere equation: 
 \begin{equation}\label{FvK:MongeAmpere}
 \det(D^2\eta)=K_0.
 \end{equation}
  Since $K_0$ is constant, equation (\ref{FvK:MongeAmpere}) can be solved by assuming $\eta$ is a quadratic function. This reduces the differential equation to an an algebraic equation.

For stretching free deformations, $E_F$ scales like $\tau^2$ and intuitively it is expected that with decreasing thickness minimizers should transition from flat to deformations that are close to being stretching free. In \cite{Gemmer2011} and \cite{Gemmer2011preprint} we made this intuitive statement precise. We showed that the global minimum of $E_F$ over stretching free deformations is obtained by a saddle shaped deformation with $\eta=xy$ and in the  vanishing thickness limit minimizers converge, up to rotations and translations in space, to a deformation with $\eta=xy$.

The one parameter family of quadratic functions $\overline{\eta}_n=y(x-\cot(\pi/n)y)$ satisfy (\ref{FvK:MongeAmpere}) as well. These functions vanish along the lines $\theta=0,\pi/n$ and thus a deformation with odd $n$-fold rotational symmetry can be constructed by taking odd periodic extensions of $\overline{\eta}_n$ about the line $\theta=\pi/n$ to form a surface $\eta_n$. These surfaces $\eta_n$ are  not smooth -- they have jump discontinuities in their second derivative across their lines of inflection -- but have finite bending energy. This is in contrast to a ridge singularity in which the bending energy diverges across the ridge \cite{Lobkovsky1996}. Note, that for $n=2$ there is no need to make such extensions.

\begin{figure}[ht]
\includegraphics[width=3.25in]{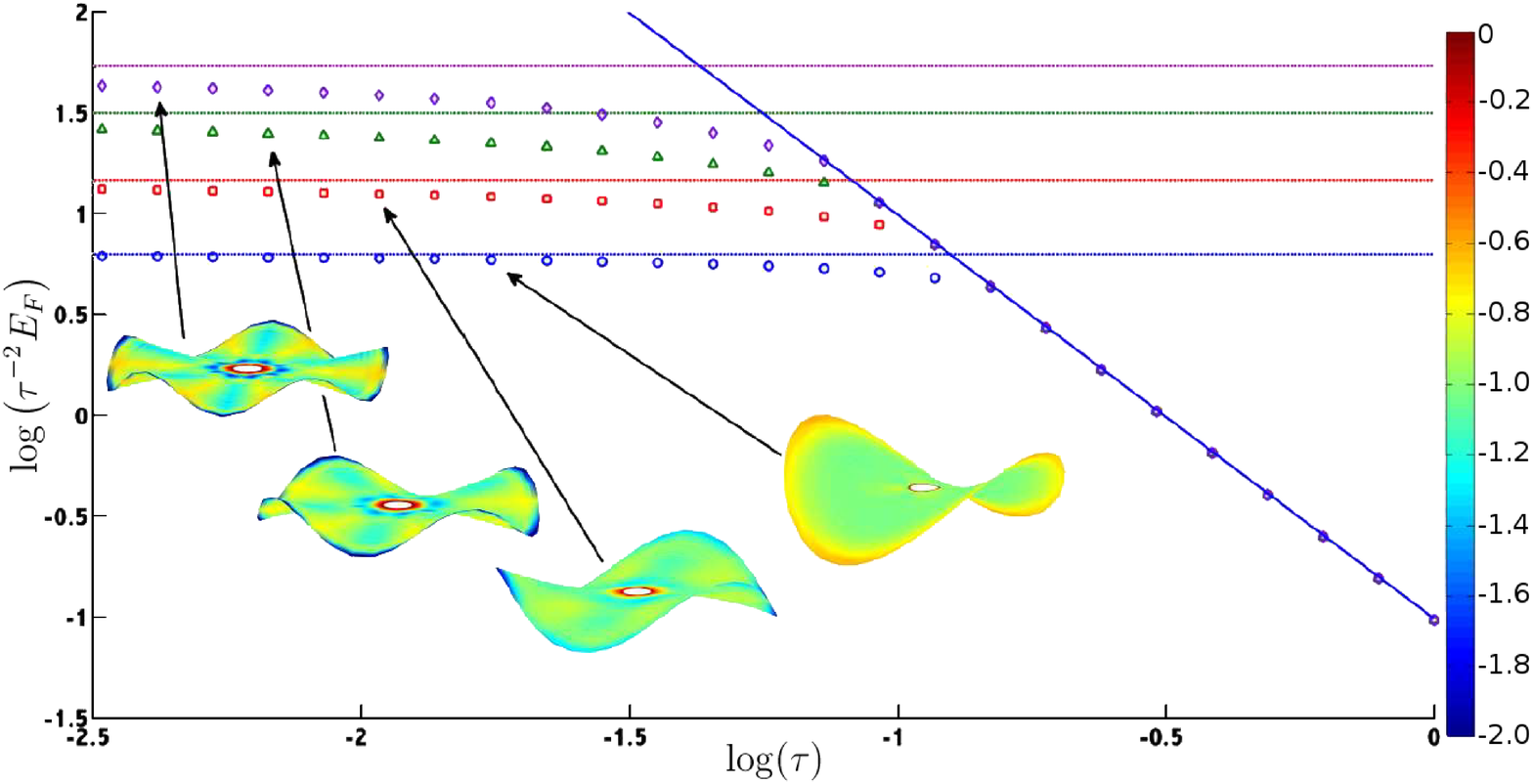}
\caption{The normalized energy of numerical minimizers of $E_F$ over deformations with $n$-fold odd rotational symmetry  for $r_0=0.1$, $R=1$, $K_0=-1$ and $\nu=1/2$. The dotted lines correspond to the energy of stretching free deformations with out-of-plane displacement $\eta_n$ while the solid line is the value $\mathcal{F}$. The configurations plotted are colored by the approximate Gaussian curvature $\det(D^2\eta)$.  \label{fig:numerics}} 
\end{figure}

The elastic energy of deformations with out-of-plane displacement  $\eta_n$ scale like $n^2\tau^2$. Furthermore,  in figure \ref{fig:numerics} it is clear that there is no regime in which higher period profiles are energetically selected over saddle shapes. Indeed, this statement was proved in \cite{Gemmer2011preprint}  by showing that minimizers with $n$ waves satisfy the following scaling for sufficiently small $\tau$:
\begin{equation}
cn\tau^2\leq \inf E_{F}[\gamma,\eta]\leq Cn^2\tau^2,
\end{equation}
where $c$ and $C$ are constants. This inequality captures the penalization paid energetically by adding more waves and proves that in the FvK model there can be no refinement of the number of waves with decreasing thickness. 

\begin{figure}[ht]
\includegraphics[width=3.2in]{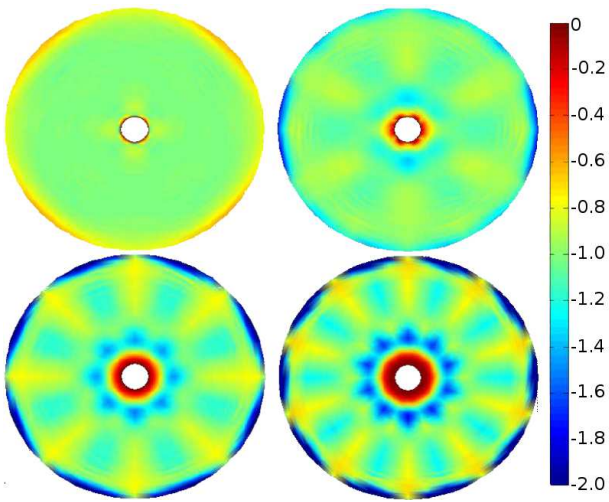}
\caption{ The annular domain colored by the Gaussian curvature of numerical minimizers over $\mathcal{A}_n$ for $\tau=0.01$, $r_0=0.1$, $R=1$, and $K_0=-1$ with $n=2,3,4,5$ clockwise from the upper left.\label{fig:boundarylayers}}
\end{figure}

Now, the exact minimizers of $E_F$ with $n$ waves are not stretching free deformations but have localized regions of stretching near their inner and outer radius and along the lines of inflection. These boundary layers are depicted in figure \ref{fig:boundarylayers}. Near the inner and outer radius of the disk these layers are the result of local stress and torque balance at the boundary. At the lines of inflection, the boundary layers are the result of corrections to the jump discontinuity in the second derivative of $\eta$. Energetically, the total energy of stretching free deformations is lowered in these boundary layers by locally balancing the bending and stretching energies to $\mathcal{O}(\tau^2)$.

We showed in \cite{Gemmer2011preprint} that near the inner and outer radius the contributions to the bending energy arising from the mean and Gaussian curvatures are both relevant. Consequently, there are two overlapping boundary layers near the edge of the annulus in which the contributions to the bending energy resulting from the mean and Gaussian curvatures are balanced with the stretching energy. The intersection of these two layers form the entire boundary layer near the edges of the disk. The width $W_K$ of the region in which the Gaussian curvature alone is reduced satisfies the following scaling:
\begin{equation} \label{FvK:BLK}
W_K(\theta)\sim t^{\frac{1}{2}}|K_0|^{-\frac{1}{4}}\csc\left(\frac{\pi}{n}\right)\left|\cos(\theta)\cos\left(\frac{\pi}{n}-\theta\right)\right|,
\end{equation} 
while for $n\geq 3$ the width $W_H(\theta)$ of the region in which the mean curvature alone is reduced scales like:
\begin{equation} \label{FvK:BLH}
W_{H}(\theta) \sim t^{\frac{1}{2}}|K_0|^{-\frac{1}{4}}\sqrt{\sin\left(\frac{\pi}{n}\right)\left|\sec(\theta)\sec\left(\frac{\pi}{n}-\theta\right)\right|}.
\end{equation}

The reason for the lack of a mean curvature boundary layer for the case $n=2$ is a result of the surface $\eta=xy$ being a minimal surface in the FvK approximation. This transition from one boundary layer in the case $n=2$ to two overlapping boundary layers for $n\geq 3$ can be detected in figure \ref{fig:boundarylayers} by noticing that within this region $K<-1$ for $n=2$ while $K>-1$ for $n\geq 3$.

For $n\geq 3$,  the width $W_I$ of the boundary layers near the lines of inflection scales like
\begin{equation}\label{FvK:BLI}
W_{I}(\rho)\sim t^{\frac{1}{3}}\rho^{\frac{1}{3}}|K_0|^{-\frac{1}{6}}.
\end{equation}
In this region the bending energy is reduced by lowering the mean curvature by regularizing the jump discontinuity in the second derivative. This boundary layer has the same scaling with thickness, but not $\rho$, for minimal ridges formed by crumpling \cite{Lobkovsky1996}. Moreover, the reduction in energy near this type of singularity is different from the regularization near a ridge singularity in which the bending energy diverges with decreasing thickness. Again for $n=2$ there is no boundary layer since  the two wave shape is smooth.

%
%

\emph{Kirchhoff approximation} -- In the Kirchhoff model the deformation $\mathbf{x}:\mathcal{D}\mapsto \mathbb{R}^3$ is assumed to be an exact isometric immersion of $\mathbf{g}$, that is a solution of the system of differential equations $(\nabla\mathbf{x})^T\cdot \nabla \mathbf{x}=\mathbf{g}$. With this ansatz the energy normalized by $t^3$ is 
\begin{equation}\label{Ki:Energy}
E_K[\mathbf{x}]=\int_{\mathcal{D}}\left[\frac{4H^2}{1-\nu}-2K\right]\,dA_{\mathbf{g}},
\end{equation} 
where $H$ and $K$ are the mean and Gaussian curvatures respectively, $\nu$ the Poisson ratio of the material, and $dA_{\mathbf{g}}$ is the area form on the surface formed by $\mathbf{x}$ \cite{Efrati2009nonEucplates, linearizedGeometry}. By Gauss's Theorema Egregium isometric immersions of $\mathbf{g}$ satisfy $K=K_0$ and thus for isometric immersions of $\mathbf{g}$ it follows that $E_{K}$  is equivalent to the energy:
\begin{equation}\label{Ki:WillmoreEnergy}
\mathcal{W}[\mathbf{x}]=\int_{\mathcal{D}}\left(k_1^2+k_2^2\right)|K_0|^{-\frac{1}{2}}\sinh(\sqrt{|K_0|}\rho)\,d\rho d\theta,
\end{equation}
where $k_1$ and $k_2$ are the principal curvatures of $\mathbf{x}$.

In this model the equilibrium configuration is selected by minimizing the full geometric bending energy in \cite{Efrati2009nonEucplates} over exact isometric immersions. Moreover, if there exist isometric immersions with finite bending energy then minimizers of the full three dimensional energy are well approximated by minimizers of $E_K$ in the vanishing thickness limit \cite{linearizedGeometry}.

Now, isometric immersions of $\mathbf{g}$ correspond to solutions of the sine-Gordon equation:
\begin{equation}\label{Ki:SineGordon}
\phi(u,v)_{uv}=|K_0|\sin(\phi(u,v)),
\end{equation}
where $u,v$ are the coordinates for a (local) parametrization by asymptotic curves and $\phi$ is the angle between the asymptotic curves \cite{gray}. Furthermore, the principal curvatures of such a parametrization satisfy $k_1^2=|K_0|\tan^2(\phi/2)$ and  $k_2^2=|K_0|\cot^2(\phi/2)$. Therefore, all of the geometric information is encoded in $\phi$ and
\begin{equation}\label{Ki:Willmore2}
\mathcal{W}[\mathbf{x}]=\int_{\mathcal{D}}\left(\tan^2\left(\frac{\phi}{2}\right)+\cot^2\left(\frac{\phi}{2}\right)\right)\,\sinh(\sqrt{|K_0|}\rho)\,d\rho\,d\theta.
\end{equation}

 Geometrically, the coordinate curves become degenerate when $\phi$ is a multiple of $\pi$ and the principal curvatures diverge. At these points the surface is not differentiable and forms sharp cusps.  Therefore, isometric immersions of $\mathbf{g}$ with finite bending energy must satisfy equation (\ref{Ki:SineGordon}) with the the constraint $0<\phi<\pi$. 

In \cite{Gemmer2011} we used a direct numerical scheme for minimizing $W$ over \emph{smooth} solutions to equation (\ref{Ki:SineGordon}). Using this scheme we provided numerical evidence that the principal curvatures of \emph{smooth isometric immersions} satisfy:
\begin{equation} \label{Ki:CurvatureGrowth}
\max\{|k_1|,|k_2|\}\geq \frac{|K_0|}{64}\exp(2\sqrt{|K_0|}R),
\end{equation}
with the surfaces realizing this bound being geodesic disks lying on hyperboloids of revolution of constant Gaussian curvature, i.e. a saddle shape. This result indicates that the bending energy of smooth isometric immersions grows exponentially with the size of the domain, a phenomenon that is not captured by the FvK model. Moreover, this result indicates that with increasing radius the principal curvatures of isometric immersions grows throughout the bulk of the domain.

Furthermore, in \cite{Gemmer2011} we showed by direct construction that there exists exact isometric immersions $A_n$ with $n$-wave profiles  which are well approximated by $\eta_n$ for small radii. The key to constructing these surfaces is the existence of a one parameter family of exact isometric immersions $A_{\theta}$, called Amsler surfaces, that contain two asymptotic lines that intersect at an angle $\theta$ \cite{Amsler}. As in the FvK approximation, if $\theta=\pi/n$ we can take the odd periodic extension of the piece of the Amsler surface bounded between the asymptotic lines to form the periodic Amsler surface $A_n$. Again, these shapes are not smooth but have jump discontinuities in their second derivatives across their lines of inflection.

\begin{figure}[ht]
\includegraphics[width=3.4in]{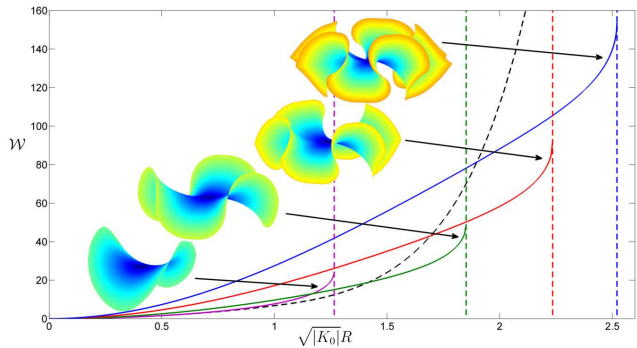}
\caption{The energy $W$ of discs cut from periodic Amsler surfaces $A_n$ with $n=2,3,4,5$ waves. The vertical dashed lines correspond to the radius where the principal curvatures of $A_n$ diverge. The dashed curve corresponds to the energy $W$ of disks cut from hyperboloids of revolution. \label{fig:amslerenergy}}
\end{figure}

Isometric immersions of geodesic balls can be ``cut'' out of $A_n$. In figure \ref{fig:amslerenergy} we plot the energy $\mathcal{W}$ of such immersions and compare their energy to those cut from hyperboloids of revolution. To be precise, there is a one parameter family of hyperboloids of revolution and for each radius we selected the hyperboloid with the lowest value of $W$.  This figure illustrates that for large values of $\epsilon=\sqrt{|K_0|}R$ the $n$-wave shapes are energetically preferred over their smooth counterparts. Additionally, in \cite{Gemmer2011} we showed that there is a critical radius $R_n\sim \ln(n)$ where the principal curvatures of $A_n$ diverge.  Beyond this radius $A_n$ no longer has finite bending energy and a profile with a higher number of waves is selected. This gives a natural mechanism for the refinement of the number of waves with increasing radius as observed in \cite{Taimina-2001}.

%
%

\emph{Discussion of results} -- From the studies in \cite{Gemmer2011} and \cite{Gemmer2011preprint} it is clear that the refinement of the number of waves with decreasing thickness observed for the gels in \cite{Shankar2011Gels} cannot be directly explained by global minimizers in the FvK and Kirchhoff approximations. Specifically, we proved in \cite{Gemmer2011} and \cite{Gemmer2011preprint} that for these gels there are only two types of global minimizers - saddles and flat sheets - in the limits of thin and thick sheets, and no intermediate asymptotic regime in which shapes with more waves have lower energy. Moreover, the bending content of the configurations in \cite{Shankar2011Gels} diverges according to the scaling $t^{-1}$ which contradicts the known existence of smooth isometric immersions of $\mathbf{g}$ for finite domains \cite{Poznyak1973}.  

But, the $n$-periodic shapes we constructed in the FvK and Kirchhoff approximations do agree qualitatively with the experimental observations in \cite{Shankar2011Gels}. Moreover, the exact isometric immersions with a periodic profile refine with increasing radius matching what is observed in \cite{Taimina-2001}.  These observations highlight the fact that the system has two relevant dimensionless numbers $t/R$ and $\epsilon = \sqrt{|K0|}R$, and the models/analysis have to account for both of these scales, instead of just the thin limit $t/R \rightarrow 0$. 

Additionally, the periodic shapes observed experimentally are the result of dynamical processes. Perhaps it is more appropriate to model this type of swelling as a gradient flow. The pattern could then be selected for dynamical reasons which could explain why local but not global minimizers appear to explain the observed shapes. The width of the boundary layers given by (\ref{FvK:BLK}-\ref{FvK:BLI}) are quantities that can be measured experimentally to determine if the local minimizers model the observed shapes.

This work was supported by the U.S. - Israel Binational Science Foundation (grant No. 2004037) and the NSF through grant No. DMS-0807501. We would also like to thank Gabrielle Meyer for providing photographs of the crocheted model of the hyperbolic plane.

\def\cprime{$'$}

\end{document}